\documentclass[amsmath,amssymb,jap,aip,reprint]{revtex4-1}

\usepackage{gensymb}
\usepackage{graphicx}

\begin{document}

\title{Calorimetric and magnetic study for Ni$_{50}$Mn$_{36}$In$_{14}$ and relative cooling power in paramagnetic inverse magnetocaloric systems}

\author{Jing-Han~Chen}
\affiliation{Department of Physics and Astronomy, Texas A\&M University, College Station, Texas 77843, USA}
\email{jhchen@tamu.edu}
\author{Nickolaus~M.~Bruno}
\affiliation{Department of Mechanical Engineering, Texas A\&M University, College Station, Texas 77843, USA}
\author{Ibrahim~Karaman}
\affiliation{Department of Mechanical Engineering, Texas A\&M University, College Station, Texas 77843, USA}
\affiliation{Department of Materials Science and Engineering, Texas A\&M University, College Station, Texas 77843, USA}
\author{Yujin~Huang}
\affiliation{School of Materials Science and Engineering, Shanghai Jiaotong University, Shanghai, 200240, China}
\author{Jianguo~Li}
\affiliation{School of Materials Science and Engineering, Shanghai Jiaotong University, Shanghai, 200240, China}
\author{Joseph~H.~Ross,~Jr.}
\affiliation{Department of Physics and Astronomy, Texas A\&M University, College Station, Texas 77843, USA}
\affiliation{Department of Materials Science and Engineering, Texas A\&M University, College Station, Texas 77843, USA}

\date{\today}

\begin{abstract}
The non-stoichiometric Heusler alloy Ni$_{50}$Mn$_{36}$In$_{14}$ undergoes a martensitic phase transformation in the vicinity of 345 K, with the high temperature austenite phase exhibiting paramagnetic rather than ferromagnetic behavior, as shown in similar alloys with lower-temperature transformations.
Suitably prepared samples are shown to exhibit a sharp transformation, a relatively small thermal hysteresis, and a large field-induced entropy change.
We analyzed the magnetocaloric behavior both through magnetization and direct field-dependent calorimetry measurements.
For measurements passing through the first-order transformation, an improved method for heat-pulse relaxation calorimetry was designed.
The results provide a firm basis for the analytic evaluation of field-induced entropy changes in related materials.
An analysis of the relative cooling power (RCP), based on the integrated field-induced entropy change and magnetizing behavior of the Mn spin system with ferromagnetic correlations, shows that a significant RCP may be obtained in these materials by tuning the magnetic and structural transformation temperatures through minor compositional changes or local order changes.

\end{abstract}


\maketitle

\section{Introduction}
Materials showing the magnetocaloric effect (MCE) have been a source of growing interest because of their 
potential for an environmentally friendly and energy efficient refrigeration technology\cite{tishin2003magnetocaloric}.
Recently, many of the off-stoichiometric Heusler alloys based on
Ni-Mn-Z (Z=In, Sb, Sn) have been reported to show a large MCE, 
across the temperature regime of the martensitic transformation\cite{:/content/aip/journal/apl/85/19/10.1063/1.1808879,krenke2005inverse,:/content/aip/journal/apl/88/12/10.1063/1.2187414,:/content/aip/journal/jap/101/5/10.1063/1.2710779,Liu2012514,Bennett201234}.
In particular, Ni-Mn-In alloys have been of interest for their large MCE͒
in the vicinity of the coupled magneto-structural first order phase transitions, and hence possible application as a working material in magnetic refrigerators\cite{PhysRevB.75.104414,:/content/aip/journal/apl/89/18/10.1063/1.2385147}.
Composition maps in the Ni-Mn-In system have been recently developed allowing the design of alloys at different working temperatures\cite{:/content/aip/journal/apl/101/22/10.1063/1.4768235}, leading to the composition studied here, Ni$_{50}$Mn$_{36}$In$_{14}$.
Contrary to typical alloys in this system, this composition has a paramagnetic rather than ferromagnetic austenite phase. 
We show that this composition exhibits a sharp transformation peak with a large field-induced entropy change at the transition temperature.

One of the common physical quantities used to compare MCE materials is the isothermal entropy change upon variation of the external magnetic field.
This quantity we have explored both by direct methods, through the integration of specific-heat data, and also by indirect methods\cite{PhysRevB.77.214439}, based on magnetization measurements.
This allows the examination of entropy contributions competing with magnetism.
However, care must be exercised in determining the specific heat close to first-order phase transitions through relaxation calorimetry.
Several groups have addressed this\cite{0953-8984-21-7-075403,Suzuki2010693,0022-3727-45-25-255001}, with one of the issues being a proper calibration of the sample puck.
In our approach to this challenge, we present a modified method which yields results in close agreement with results obtained through indirect magnetic measurements.

As further comparison of the MCE, we examine the relative cooling power (RCP)\cite{franco2012magnetocaloric} which 
represents the amount of transferred heat between the hot and cold reservoirs under a magnetic refrigeration cycle\cite{Pecharsky199944}.
Values extracted from direct and indirect methods provide a compelling demonstration that the corresponding entropies are in very close agreement.
The RCP for the present material is smaller than comparable materials operating at lower temperatures, which we show results from the separation between the austenite Curie temperature ($T_c$) and the operating temperature.
However, further analysis shows that a significant RCP would be achievable in such systems by suitable tuning of structural and magnetic transition temperatures.

\section{Experimental Procedures}
\subsection{Sample Preparation}
Bulk polycrystalline Ni$_{50}$Mn$_{36}$In$_{14}$ (nom. at. $\%$) alloys were prepared using arc melting in a protective argon atmosphere from $99.9\%$ pure constituents.
Ingots were flipped and re-melted three times to promote homogeneity. The resulting 35g button was cut into plates using a high speed diamond wafering blade.
The samples were then quenched in ice water after they were homogenized at 900$\celsius$ for 24 hours in a quartz vial under a protective argon atmosphere to prevent oxidation.
One of these plates was used for $M$-$T$ measurements, and another identically treated plate was used for both $M$-$H$ and specific heat measurements.

\begin{figure}
\includegraphics[width=0.6\columnwidth]{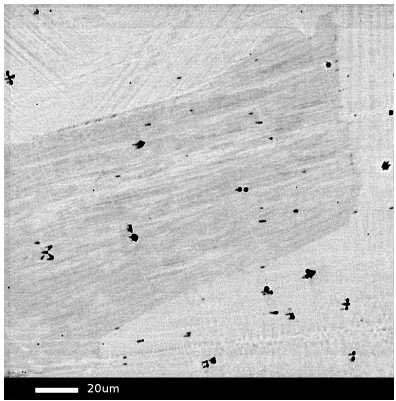}
\caption{Back-scattered electron image of the sample Ni$_{50}$Mn$_{36}$In$_{14}$ after heat treatment (scale bar $20\mu m$ in the bottom of the figure). The very dark regions are empty cavities and image contrast shows grain size.}
\label{wds}
\end{figure}

Electron microprobe measurements were carried out using wavelength dispersive spectroscopy
 methods on a Cameca SX50, equipped with four
 wavelength-dispersive x-ray spectrometers. The final composition Ni$_{49.54}$Mn$_{36.12}$In$_{14.34}$ was found to be close to the target composition.
The samples after the heat treatment were found to be homogeneous by back-scattered electron imaging (shown in Fig.~\ref{wds}) and microprobe analysis.

\subsection{Magnetic Measurements\label{magexp}}
\begin{figure}
\includegraphics[width=\columnwidth]{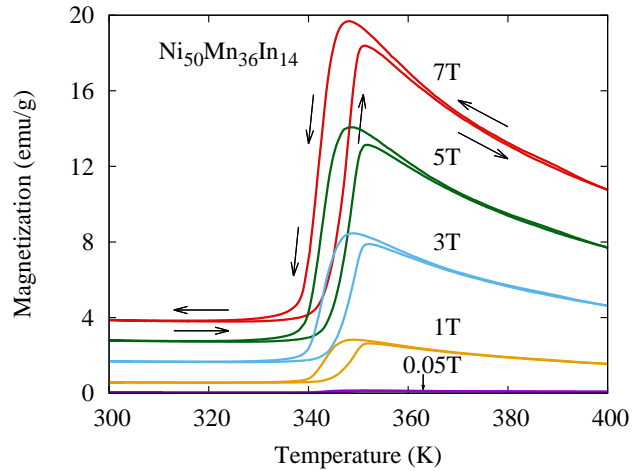}
\caption{Temperature dependence of the magnetization for the Ni$_{50}$Mn$_{36}$In$_{14}$ sample at different fields (0.05, 1, 3, 5 and 7 T as shown).}
\label{mt}
\end{figure}

Iso-field magnetization measurements were carried out using a Quantum Design (QD) Magnetic Property Measurement System on a sample from the center of the homogenized sample plate.
Figure~\ref{mt} shows the measurement results.
The transition curves as shown in Fig.~\ref{mt} correspond to the well-known martensitic transformation. 
The results indicate a transition from a paramagnetic austenite to an anti-ferromagnetic (or similar low-moment) martensite upon cooling in the vicinity of 345 K.
The forward (austenite to martensite) transformation is observed on cooling the samples and the reverse transformation (martensite to austenite) is observed on sample heating. 
Outside of the transformation hysteresis region, the transition is nominally complete.
As shown in Fig.~\ref{mt}, the transition temperatures slightly shift to lower temperatures as the field increases.

\begin{figure}
\includegraphics[width=\columnwidth]{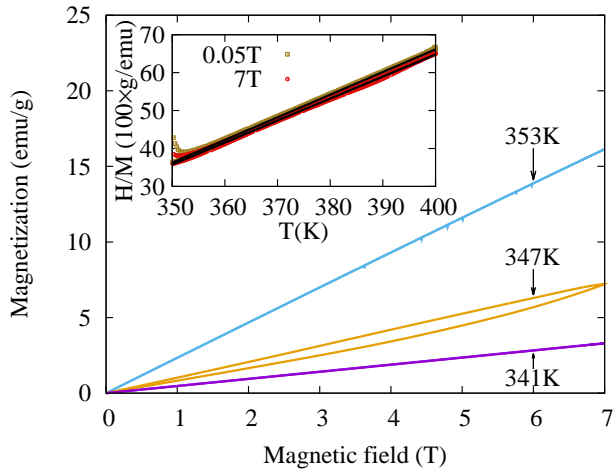}
\caption{Three representative isothermal magnetization measurements. Magnetic hysteresis was only observed between 341 and 353 K (including the 347 K trace shown).
		Inset: Iso-field measurements for 0.05 T and 7 T, along with Curie-Weiss fittings (black lines) showing strong linearity of the results in the paramagnetic austenite phase.} 
\label{mh}
\end{figure}

The Curie-Weiss law,
\begin{equation}
M=\frac{N_A}{3k_B}\mu_{eff}^2\frac{H}{T-T_c},
\end{equation}
where $\mu_{eff}=g\mu_B\sqrt{J(J+1)}$, $T_c$ is the Curie temperature and $g=2$,
was used to fit the high temperature ($>$ 360 K) magnetization curves. 
Fitting in all five fields in Fig.~\ref{mt} yielded nearly identical $T_c$ values in the range $289\pm0.5$ K (fitting uncertainty $0.03-0.04$), and spin very close to $J=2.0$ per Mn (range 2.01 to 2.04, fitting errors $< 0.004$).
This fitted spin value was obtained by assuming that the density of magnetic moments is identical to the manganese ion density.
The fitted curves are shown in the inset of Fig.~\ref{mh} for two of these fields.
The linearity of these curves in $M$ and $T$ demonstrates the paramagnetism of the austenite phase.
These results with a local magnetic moment of $4\mu_B$ agree with reported studies\cite{Li201335,0953-8984-21-23-233201,PhysRevB.86.214205,:/content/aip/journal/apl/97/24/10.1063/1.3525168} and are consistent with magnetic moments in the sample attributed to manganese with at most a very small moment on other atoms.
Note that the Curie temperature apparent here is very similar to $T_c\approx293$ K observed in Ni$_{50}$Mn$_{34}$In$_{16}$\cite{:/content/aip/journal/apl/85/19/10.1063/1.1808879}.
Even though the structural transition temperatures are in general very sensitive to composition in the Ni-Mn-In alloys, $T_c$ of the austenite is nearly composition-independent, a feature that holds even for the present case where the ferromagnetic fluctuations in the austenite phase are interrupted by the higher-temperature structural transition.

Besides increasing/decreasing the temperature of the sample, changing the external magnetic field can also drive the magneto-structural transition.
However, for cases including a first-order transition through scanning the magnetic field, incorrect probing protocols can lead to spurious magnetic entropy changes\cite{Caron20093559,Bruno201466,franco2012magnetocaloric,zhang2008spike,tocado2009entropy,carvalho2011isothermal}.
In order to correctly probe the phase transition, the following isothermal measurement was performed, resulting in curves such as plotted in Fig.~\ref{mh} (main plot).
The $M$-$H$ measurements at a sequence of temperatures always included a loop bringing the sample below the lowest temperature, 325 K.
The magnetization was recorded at each temperature with the field increasing to the maximum value and also while it was brought back to zero.
Next, the temperature was reduced, bringing the sample into the complete martensite region, before going to the next measurement temperature.
This was done with steps of $\Delta T=+3$ K near the phase transition, with the loop in temperature always performed at zero field before every isotherm.
The magnetization for temperatures higher than 353 K and lower than 341 K shows similar paramagnetic behavior with no observable hysteresis or nonlinearity
so only three representative curves are shown in Fig.~\ref{mh}.

\subsection{Specific Heat Measurements\label{hcexp}}
Specific-heat measurements were performed using a Physical Property Measurement System (PPMS) from QD.
Samples were affixed using a thin layer of grease to the small sample platform of a standard puck, 
with the platform connected by thin wires to the body of the puck. 
For temperatures away from the structural transition, the so-called 2-$\tau$ method was 
used, in which the heating and cooling curves are fitted following a small heat pulse.
This analysis is based on a model developed by Hwang\cite{:/content/aip/journal/rsi/68/1/10.1063/1.1147722} in which the heat transfer between the platform and the sample is explicitly taken into account. 
In determining the sample's specific heat ($C_p$), the relationship governing the heating and cooling curves is, 
\begin{equation}
C_{total}(T)=\frac{-K_w(T)(T-T_b)+P}{dT/dt},
\label{heat_flow}
\end{equation}
where $C_{total}$ is the total specific heat including both $C_p$ due to the sample and the contribution of the platform, $P$ is the known heat pulse power, $T$ is the sample temperature measured by the platform thermometer, $T_b$ is the separately measured puck temperature, and $K_w$ is the thermal conductance from the platform, typically dominated by the support wire conductance.
Note that $P$ goes to zero during cooling while $dT/dt$ changes to a negative sign.
Assuming $C_{total}$ and $K_w$ remain the same upon heating and cooling, these parameters can be obtained by analyzing the heating and cooling curves together.	
However, this method is invalid when the specific heat involves hysteresis, such as is typical for a first order transition.
Also if a long heat pulse is used to drive the transition to completion, $(T-T_b)$ can become large. 
This can invalidate the standard calibration of $K_w$ for which $(T-T_b)$ is small and the wires are held at a nearly uniform temperature.

These issues have been addressed in various ways\cite{0953-8984-21-7-075403,Suzuki2010693,0022-3727-45-25-255001}, including models introduced to consider changes in the wire conductance and also
changes in the base temperature during a long heat pulse.
In our study, to probe the phase transition region we recorded changes in temperature during a long
heat pulse, turned on in the temperature range of complete martensite
and turned off when complete austenite was achieved.
This requires optimizing the pulse settings by trial and error to cover the
transition region, and long wait times to increase the base temperature stability.
The raw temperature vs. time traces were obtained from the system controller, and analyzed separately.
From measurements away from the hysteresis region using the 2-$\tau$ measurement scheme, we determined that the thermal conductance of the grease
exceeds that of the wires by a factor of at least $100$ in all cases. This ensured that the temperature
discrepancy between the platform and sample is negligible.

\begin{figure}
\includegraphics[width=\columnwidth]{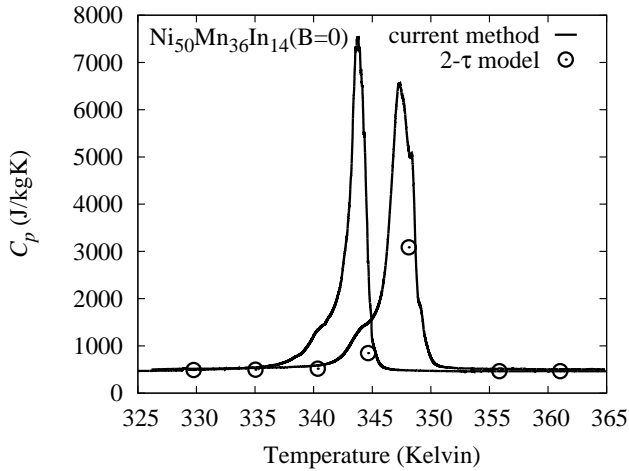}
\caption{Solid curves: specific heat obtained from heating and cooling measurements analyzed as described in text. 
		Open symbols: results from 2-$\tau$ model, with small heat pulses giving $\Delta T/T = 1\%$, showing good agreement outside of first order transition regime.}
\label{hysteresis}
\end{figure}

In order to calibrate $K_w$ for the conditions of large $\Delta T $($=T-T_b$) in eq.~\ref{heat_flow}, we made use of the lack of hysteresis outside of the transition, as known physically and observed here.
Because of this feature, for heating-cooling cycles bracketing the transition, the 2-$\tau$ model remains valid for extracted $K_w$ values at the large-$\Delta T$.
At the small-$\Delta T$ region, the conditions of the standard calibration procedure are reproduced, so the standard calibration provided these values.
In our study, a modified wire conductance $K_w^\prime$ is used
\begin{equation}
K_w^\prime(T)=K_w((1-x)T_b+xT),
\end{equation}
where $K_w(T)$ is obtained from the standard calibration, and a single extrapolation parameter $x$ was adjusted until the heating and cooling curves became consistent for temperatures above the transition.
Finally, from the time recordings of temperature during and after the heat pulse, we obtained the heating and cooling specific heats by utilizing eq.~\ref{heat_flow} and $K_w^\prime$.
This procedure yields a dense set of data for $C_p$ in the transition region, 
	 characterized by a large peak for the first-order transition, 
	 and wings that connect smoothly to data obtained by the standard 2-$\tau$ model covering a wide temperature range. 
This is shown in Fig.~\ref{hysteresis}, where the 2-$\tau$ model data (not valid near the transition) are given by isolated symbols.

\begin{figure}
\includegraphics[width=\columnwidth]{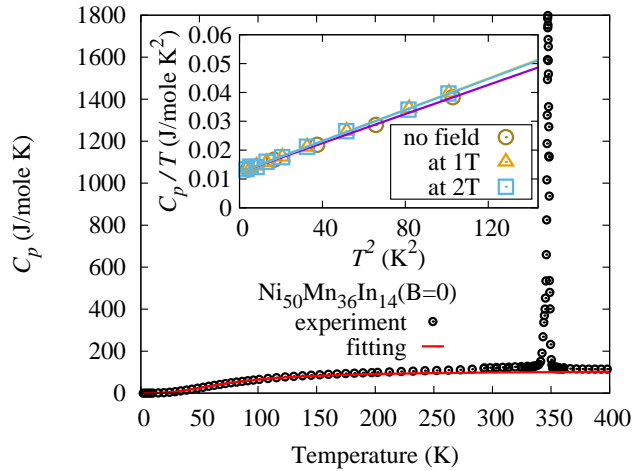}
\caption{Full-range specific heat, combining heating-curve results bracketing the martensitic transformation, with small-pulse results outside the hysteresis region. 
		Solid curve: combined electronic and Debye-approximation contribution with $\gamma= 0.012$ J/mole K$^2$ and $\theta_D = 320$~K.
		Inset: low temperature experimental data for three fields as indicated plotted along with fitted results (solid lines) for $\gamma$, found to be field independent.}
\label{gamma-inset}
\end{figure}

Fig.~\ref{gamma-inset} shows the results of this method for the Ni$_{50}$Mn$_{36}$In$_{14}$ sample 
in the range of the martensitic transformation, combined with 2-$\tau$ analysis for the remainder of the temperature range. 
Data for the transition region are from the temperature-rise portion of the measurement, and include the superposed results of several trials. Results are given pre mole of Ni$_2$Mn$_{1.44}$In$_{0.56}$. 

\begin{figure}
\includegraphics[width=\columnwidth]{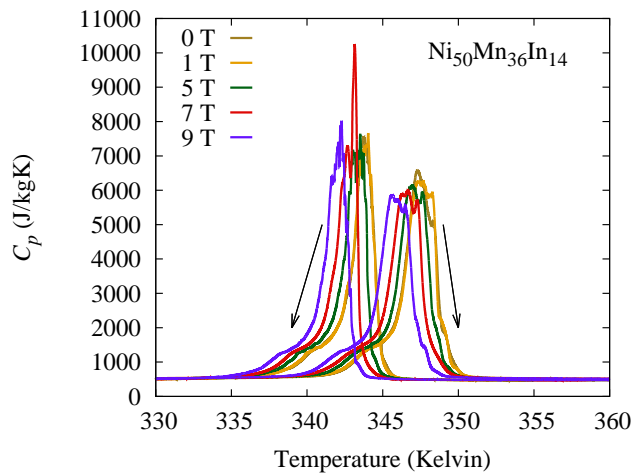}
\caption{Heating curve and cooling curve specific heat results obtained under various applied magnetic fields (0, 1, 5, 7 and 9 T) as shown.}
\label{field}
\end{figure}

Specific heat measurements were also performed across the transition under external magnetic fields of 1, 5, 7, and 9 T, with the results shown in Fig.~\ref{field}.
The observed structure at the peak is due in part to the narrow transformation in this composition, and the small sampling interval for the finite derivative, particularly on cooling.
However, the integrated quantities are quite consistent, as shown below.
The first order transition is shifted to lower temperatures as the external field increases, in agreement with the magnetic measurements under corresponding fields in Fig.~\ref{mt}.

\section{Specific-Heat Contributions Analysis}

Owing to the diffusionless character of the martensitic transformation,
the configurational contributions to the entropy change are believed to be absent\cite{:/content/aip/journal/jap/93/10/10.1063/1.1556218}, which simplifies the sources of the specific heat. 
The total specific heat for this system should be thus described by three main contributions, which are the electronic, vibrational and magnetic parts:
\begin{equation}
C_p\approx C_{el}+C_{vib}+C_{mag}.
\label{ctotal}
\end{equation}
The electronic specific heat contribution can be described by the known linear-$T$ behavior,\cite{mizutani2001introduction}
\begin{equation}
C_{el}(T)\cong\left(\frac{\pi^2}{3}\right)k_B^2N(E_F)T=\gamma T,
\label{cel}
\end{equation}
where $N(E_F)$ is the density of states at the Fermi energy.
The vibrational contribution\cite{mizutani2001introduction} based on a Debye model is
\begin{equation}
C_{vib}(T)\cong 9Nk_B\left(\frac{\theta_D}{T}\right)^3\int_0^{\theta_D/T}\frac{x^4e^x}{(e^x-1)^2}dx
\label{cvib}
\end{equation}
where $N$ is the total number of atoms and $\theta_D$ is the Debye temperature, which we treated as a constant.
Note that at the transition temperature the vibrational term is quite close to the classical limit,
 considerably reducing the dependence on $\theta_D$, in the limit that the harmonic approximation remains valid.
Since the magnetic contribution is expected to be small in the low temperature limit\cite{Chernenko2008e156}
we fitted to $C_p/T$ vs. $T^2$ at low temperatures (inset to Fig.~\ref{gamma-inset}), yielding
$\gamma = 0.0124(2)$ J/mole K$^2$, independent of the field within the experimental uncertainty.
Furthermore $\theta_D= 320$ K was obtained by fitting the data up to 100 K.
A calculation using these two values, extended to 400 K, is given by the solid line in Fig.~\ref{gamma-inset}
and shows good agreement with the experimental data (solid circles) in the same figure.
As can be seen, $C_p$ greatly exceeds the extrapolated values of $C_{el}$ and $C_{vib}$ near the transition, with the difference representing $C_{mag}$ to the extent that eq.~\ref{cel} is valid.

\section{Entropy Change and Relative Cooling Power Analysis}

The MCE is intrinsic to magnetic solids and is induced via the coupling of the magnetic sublattice with the magnetic field\cite{:/content/aip/journal/jap/86/1/10.1063/1.370767}.
Accurate determination for the MCE can be obtained from both indirectly via magnetization and directly through specific heat. 
Both methods will be discussed and compared in the following section.

\subsection{Magnetocaloric Effect from Magnetic Measurement}

The most common indirect experimental determination of field-induced entropy change is obtained from the 
isothermal magnetization by scanning the magnetic field over a series of closely spaced temperature intervals\cite{Willard2013173,:/content/aip/journal/jap/86/1/10.1063/1.370767,0953-8984-21-23-233201}.
Representative data are shown in Fig.~\ref{mh} and the probing protocol has been described in section~\ref{magexp}.
Because the transition is first order, the Maxwell relation does not apply and instead should be replaced by a Clausius-Clapeyron method, as has been confirmed by experiments\cite{PhysRevLett.83.2262}.

This isothermal field-induced entropy can be obtained from the magnetization by the relation\cite{Amaral20101552}
\begin{equation}
\begin{split}
&\Delta S(T, 0\to H)=\mu_0\frac{\partial}{\partial T}\left(\int_0^H M(T,H^\prime)dH^\prime\right)\\
&\cong\frac{\mu_0}{\Delta T}\left[\int_0^HM(T+\Delta T,H^\prime)dH^\prime-\int_0^HM(T,H^\prime)dH^\prime\right].
\end{split}
\label{mhentropyeq}
\end{equation}
Results obtained from this expression can be found in Fig.~\ref{entropy} (in symbols), for the fields of 1, 5 and 7 T, which correspond to specific heat measurements.

It is important to recognize that the magnetic MCE measurements may be intrinsically
erroneous if changes in the material are time dependent, i.e., if the response is not instantaneous, because the experiments were carried out in a sweeping magnetic field. 
In our experiment, we lowered the sweeping speed to a rate ($=25$ Oe/s) for which the measurement results were found to be consistent with data for a higher speed, in order to avoid this situation.

\subsection{Magnetocaloric Effect from Calorimetric Measurements}

As mentioned in the end of the previous section, another way to characterize MCE is directly from the calorimetric measurements.
The specific heat measured at constant pressure as a
function of temperature in constant magnetic field $C_p(T,B)$
provides the most complete characterization of MCE materials 
since the total entropy change of the MCE material can be calculated
from the specific heat by the relation\cite{PhysRevB.77.214439}
\begin{equation}
\Delta S(T, 0\to H)=\int_0^T\frac{C_p(T^\prime,H)-C_p(T^\prime,H=0)}{T^\prime}dT^\prime.
\label{cpentropyeq}
\end{equation}

\begin{figure}
\includegraphics[width=\columnwidth]{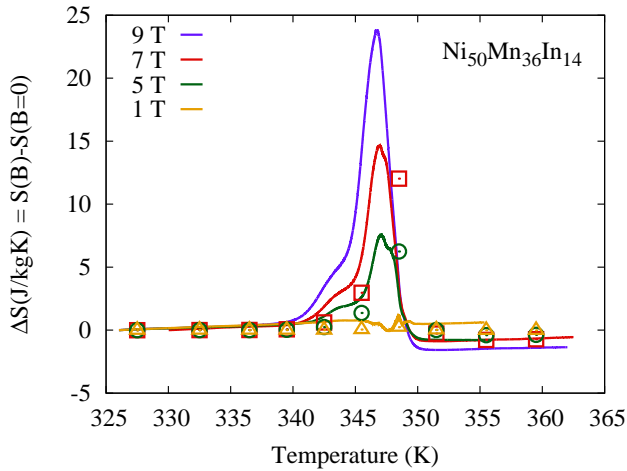}
\caption{Reverse (martensite to austenite) field-induced entropy difference ($S(T,H)-S(T,0)$) based on the heating-curve specific heat results
		(solid curve) along with entropy results from the corresponding magnetic analysis (symbols).}
\label{entropy}
\end{figure}

The field-induced entropy change obtained from the specific heat measurements while heating in various fields (Fig.~\ref{field}) is shown in Fig.~\ref{entropy}.
The results from magnetic measurements plotted in the same figure appear to be in good agreement.
A similar comparison was reported recently for Co-doped Ni-Mn-In materials\cite{0022-3727-45-25-255001}.
This implies that the isothermal entropy change due to magnetic properties dominates through the martensitic transformation.
Although the other possible physical mechanisms (such as, crystal structure or phonon) can not be completely ruled out, their contributions are clearly relatively minor.

\subsection{Relative Cooling Power}

Another relevant quantity for evaluating the performance of MCE materials is the amount of transferred heat
between cold and hot reservoirs in an ideal refrigeration cycle\cite{:/content/aip/journal/jap/90/9/10.1063/1.1405836,gschneidner1999recent}.
This is referred to as the relative cooling power (RCP),
\begin{equation}
\text{RCP}(H)=\int_{T_{cold}}^{T_{hot}}\Delta S(T, 0\to H)dT,
\label{rcp}
\end{equation}
where $T_{cold}$ and $T_{hot}$ are the temperatures of the two reservoirs.
Therefore we can obtain RCP by calculating the area under the $\Delta S$ curves.
Eq.~\ref{rcp} can also be rewritten\cite{Amaral20101552} by using eq.~\ref{mhentropyeq}
\begin{equation}
\begin{split}
\text{RCP}(H)&=\int_{T_{cold}}^{T_{hot}}\Delta S(T, 0\to H)dT\\
&=\int_0^H M(T_{hot},H^\prime) dH^\prime-\int_0^H M(T_{cold},H^\prime) dH^\prime
\end{split}
\label{rcpmh}
\end{equation}
which indicates that RCP can be determined from isothermal magnetic measurements at only two temperatures without knowing the details of the magnetic entropy at points between.
Thus, despite the very sharp transition in this case, a larger number of magnetic isotherms (including cooling loops below the transformation) was not required.
On the other hand, since the entropy change obtained from specific heat data is much denser, we could calculate the entropy integral directly from the data.
The results provide an additional useful comparison between methods.

\begin{figure}
\includegraphics[width=\columnwidth]{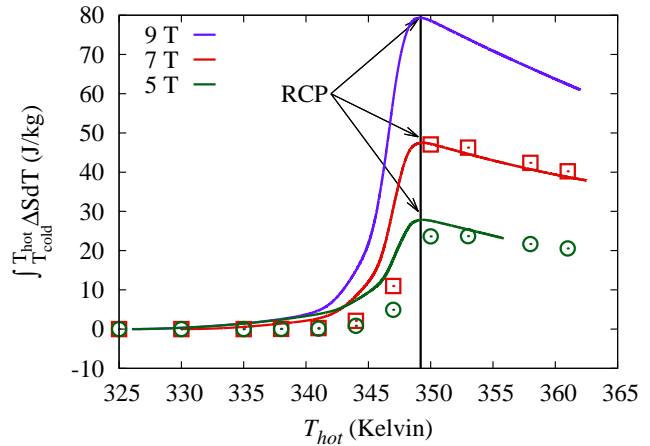}
\caption{Integrated entropy obtained from specific heat (solid curves) along with RCP for complete transformation, compared to values 
		obtained from the difference of isothermal magnetic measurements (symbols).}
\label{rcp_mag}
\end{figure}

Generally, the RCP can be considered to be given by eq.~\ref{rcp} or eq.~\ref{rcpmh} evaluated with $T_{hot}$ and $T_{cold}$ bracketing the transition, corresponding to full martensite to austenite conversion. 
However, Fig.~\ref{rcp_mag} displays the corresponding integrations evaluated with $T_{hot}$ allowed to vary across the transition.
This provides an excellent comparison between the results obtained by both methods, since as eq.~\ref{rcpmh} shows the structure of $\Delta S(T,0 \to H)$ is properly accounted for even at intermediate temperatures where $M(H)$ has not been measured. 
The 1 T entropy change is small, giving an RCP integral (not shown) indistinguishable from the background. 
However, in 5 T and 7 T the results are in very good agreement, including the observed asymmetry of the $\Delta S$ peaks, which is reproduced with both methods.
Also note that above 350 K, the $\Delta S$ integration curves slope downward since here in the paramagnetic phase, a conventional (non-inverse) MCE sets in.

To define the RCP, in practice we chose the point where the integration curves of Fig.~\ref{rcp_mag} reach a maximum, which is 349 K for all cases, at the temperature indicated in the figure.
The resulting RCP values are 28 J/kg in 5 T, 48 J/kg in 7 T, and 79 J/kg in 9 T. 
For comparison RCP = 242 J/kg in 7 T has been reported for an annealed NiCoMnSn ribbon\cite{Bruno201466}, and 104 J/kg in 5 T for bulk Ni$_{50}$Mn$_{34}$In$_{16}$\cite{0022-3727-40-7-005}.
Therefore, in the present case even though a large field-induced entropy change is observed, the RCP remains smaller than can be achieved in corresponding materials exhibiting a ferromagnetic austenite phase.

\subsection{RCP in paramagnetic systems}
We now consider more generally the RCP obtainable in similar alloy systems with the transformation temperature $T_m$ closer to $T_c$, where the austenite paramagnetic response will be stronger.
As we established, the Ni$_{50}$Mn$_{36}$In$_{14}$ austenite is described very accurately as Curie-law paramagnetic with each Mn having a spin fitted to $J=2.0$, and ferromagnetic correlations corresponding to $T_c=289$ K.
Similar behavior is observed in other Ni-Mn-In systems, and as noted above $T_c$ is found to be relatively independent of composition.
For the present alloy, since the magnetization curves are linear close to the transition in both phases (Fig.~\ref{mh}), eq.~\ref{rcpmh} can be written, RCP$(H)=\Delta\chi H^2/2$, with $\chi=M/H$ being the susceptibility.
Based on the observed small martensite magnetic response, its contribution to RCP in 9 T is $-\chi_mH^2/2= -21$ J/kg, which decreases rapidly at lower fields. 
Closer to $T_c$, the paramagnetic response of the austenite will become nonlinear, and to address this situation we have adopted a mean field approach\cite{blundell2001magnetism}.
Except for the critical region very close to $T_c$ in small fields, this gives a good approximation for the spin polarization, and as long as $T$ remains above $T_c$, coercivity effects need not be considered. 

\begin{figure}
\includegraphics[width=\columnwidth]{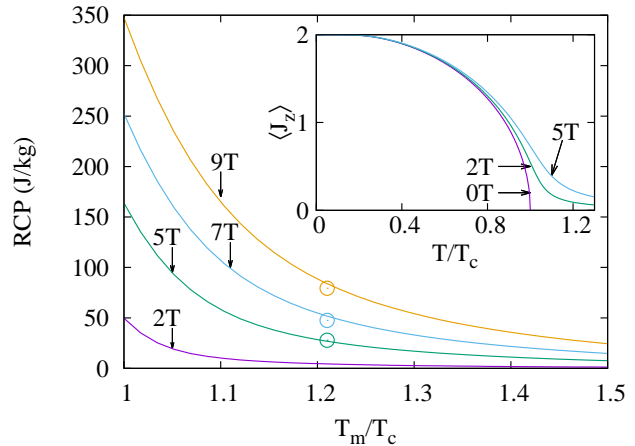}
\caption{Relative cooling power vs. martensitic transformation temperature calculated for Ni-Mn-In compositions with paramagnetic austenite phase, as described in text.
		Inset: computed austenite average spin moment for indicated fields.}
\label{rcpplot}
\end{figure}

The inset of Fig.~\ref{rcpplot} shows representative spin polarization results obtained following a standard derivation\cite{blundell2001magnetism}, with the mean spin following a Brillouin function containing the applied magnetic field enhanced by the local exchange field.
We assumed each Mn to have a local moment with $J=2$ and $g=2$, coinciding with the magnetization fit, and used $T_c=289$ K as measured, which determines the size of the exchange field.
The magnetization is obtained from the average spin using the known moment per Mn, along with the Mn-ion density per kg for Ni$_{50}$Mn$_{36}$In$_{14}$.
No adjustable parameters are needed to obtain $\int MdH$, which is the integral appearing in eq.~\ref{rcpmh}. 
Fig.~\ref{rcpplot} shows RCP results obtained this way vs. $T_m/T_c$, where the temperature at which the integration is evaluated is assumed to match $T_m$ for different compositions along the horizontal axis.
In this calculation we also assumed that the martensite-phase contribution to RCP, $-\chi_mH^2/2$, remains unchanged. 
While in practice this term will vary with processing, note that this term is in any case small, particularly near $T_c$.
The circles on the graph represent the RCP extracted from specific heat for Ni$_{50}$Mn$_{36}$In$_{14}$ (with $T_m/T_c = 1.21$).
These are in good agreement with the calculated curves, including the 9 T value.

More generally, Fig.~\ref{rcpplot} shows the RCP that should be attainable in compositions similar to Ni$_{50}$Mn$_{36}$In$_{14}$ if $T_m$ is adjusted along the horizontal axis through composition or local ordering changes.
This includes magnetic-only contributions to the entropy, shown here to be nearly identical to the total entropy change. 
For $T_m$ close to $T_c$ the RCP values become comparable to those reported in other Ni-Mn-X materials, as described above.
Note that an approach typically used for these materials has been to maximize the austenite magnetization, for example\cite{liu2012giant} by substitution of Co, in order to enhance the RCP. 
However, these results predict that compositions with $T_m\sim Tc$ having small or zero spontaneous magnetization in both phases may still exhibit comparable results. 
Since magnetic coercivity effects will not contribute to the thermal hysteresis, optimizing materials of this type may represent a promising approach for obtaining useful MCE materials.

\section{Summary}
In this study, we examined the magnetocaloric effect in Ni$_{50}$Mn$_{36}$In$_{14}$, heat treated to exhibit a narrow structural phase transition to the martensitic phase near 345 K.
Above the transformation this composition is paramagnetic with no spontaneous magnetization, unlike similar compositions typically studied with somewhat lower structural transition temperatures.
Field-induced entropy changes were analyzed both through indirect magnetic and direct calorimetric measurements.
For the latter, we demonstrated a procedure based on relaxation calorimetry using extended heating and cooling curves extending across the first-order phase transition.
The results are in excellent agreement with magnetization-based methods, which provide a firm basis for the analytic evaluation of field-induced entropy changes.
In particular, the relative cooling power (RCP) can be assessed in this way, and the associated analysis shows that a large RCP may be generated in materials with the structural transition tuned very close to the Curie temperature.

\begin{acknowledgments}
This material is based upon work supported by the National Science Foundation under Grant No. DMR-1108396, and by the Robert A. Welch Foundation (Grant No. A-1526). We gratefully acknowledge assistance with PPMS measurement procedures from Neil Dilley and Randy Black at Quantum Design.
\end{acknowledgments}

\bibliography{aipsamp}

\end{document}